\documentclass[fleqn,10pt]{wlscirep}
\usepackage[utf8]{inputenc}
\usepackage[T1]{fontenc}
\usepackage{lineno}

\title{A Dataset of the Operating Station Heat Rate for 806 Indian Coal Plant Units using Machine Learning}

\author[1]{Yifu Ding}
\author[2]{Jansen Wong}
\author[3]{Serena Patel}
\author[4]{Dharik Mallapragada}
\author[1]{Guiyan Zang}
\author[1]{Robert Stoner}
\affil[1]{MIT Energy Initiative, Massachusetts Institute of Technology}
\affil[2]{Department of Electrical Engineering and Computer Science, Massachusetts Institute of Technology}
\affil[3]{The Brattle Group}
\affil[4]{Department of Chemical and Biomolecular engineering, Tandon School of Engineering, New York University}

\begin{abstract}
India aims to achieve net-zero emissions by 2070 and has set an ambitious target of 500 GW of renewable power generation capacity by 2030. Coal plants currently contribute to more than 60\% of India's electricity generation in 2022. Upgrading and decarbonizing high-emission coal plants became a pressing energy issue. A key technical parameter for coal plants is the operating station heat rate (SHR), which represents the thermal efficiency of a coal plant. Yet, the operating SHR of Indian coal plants varies and is not comprehensively documented. This study extends from several existing databases and creates an SHR dataset for 806 Indian coal plant units using machine learning (ML), presenting the most comprehensive coverage to date. Additionally, it incorporates  environmental factors such as water stress risk and coal prices as prediction features to improve accuracy. This dataset, easily downloadable from our visualization platform, could inform energy and environmental policies for India's coal power generation as the country transitions towards its renewable energy targets.

\end{abstract}
\begin{document}

\flushbottom
\maketitle

\thispagestyle{empty}

\section*{Background \& Summary}

Under economic development and rapid population growth, India's electricity demand is projected to double by 2040 \cite{iea_india_2021}. The current Indian power system depends heavily on high-emission coal plants. The installed power capacity of coal plants reached 205 GW and accounted for 49\% of the total capacity in 2022 \cite{central_electricity_authority_installed_2022}. These coal plants contribute more than 60\% electricity generation  \cite{central_electricity_authority_national_2022}. India aims to achieve the net-zero emissions by 2070 and is on their way to achieving their target of 500 GW of installed renewable energy capacity by 2030 \cite{ministry_of_power_government_of_india_india_2022}. Despite these targets, India has not proposed a comprehensive phase-out plan for coal plants. Instead, the government still permits new construction, life extension, and renovation of the aging coal plants \cite{ministry_of_power_phasing_2023}. 

Historically, Indian coal plants have low thermal efficiency, mainly due to the usage of low-quality domestic coal and low-efficiency boilers \cite{mallapragada_life_2019}. More than 50\% of Indian coal plants still use sub-critical boiler technology, and the first super-critical coal plant with a higher thermal efficiency was commissioned in 2012 \cite{ns_energy_kudgi_2012}. As a result, coal power generation in India produces significant carbon dioxide emissions and air pollution, contributing to many premature deaths \cite{cropper_mortality_2021}. In addition, most of India’s thermal power generation depends on freshwater cooling, which exacerbates water stress issues and leads to water shortage related operational disruptions \cite{tianyi_luo_parched_2018}. Upgrading and decarbonizing coal power generation became a near-term pressing energy issue in India. 

\subsection*{Data Gap in the Existing Databases of Indian Coal Plants}

While the government sector and research institutes have researched on Indian power systems using power dispatch \cite{sengupta_subnational_2022} and capacity expansion models \cite{deshmukh_least-cost_2021, rudnick_decarbonization_2022, central_electricity_authority_draft_2023}, their methodologies to characterize coal plants often oversimplify plant thermal efficiencies. The SHR, defined as the ratio of the heat input to the station to the electricity generated, represents the plant thermal efficiency to calculate the fuel consumption and power output. A high SHR value indicates a low thermal efficiency of coal-fired power unit. Yet, many studies characterise SHR using a single value \cite{deshmukh_least-cost_2021, central_electricity_authority_draft_2023, rudnick_decarbonization_2022} or a fitting curve \cite{sengupta_subnational_2022} across the entire fleet. In practice, the operating SHR may vary significantly depending on the plant design, ambient conditions, and operating regimes. These approximations can lead to the under-estimation of coal consumption and carbon emission, which results in ineffective policy implications -- a crucial concern given coal's dominance in India's energy system. A more granular, plant-specific characterization of the operating SHR for the entire Indian coal plant fleet is necessary for accurate modeling and effective policy making.  

 Table \ref{Databases} shows the existing databases of India coal plants with the SHR characterization methods used. These datasets are incomplete in different aspects. The first database is from the Global Energy Monitor (GEM) (\url{https://globalenergymonitor.org/projects/global-coal-plant-tracker/}) which records 840 Indian operating coal plant units to date \cite{global_energy_monitor_global_2024}. It also includes the location, the commission year, and the boiler types of each coal-fired power unit, but the SHR and emission factors are based on linear estimations which could greatly deviate from the actual values. The second and third databases are from the Central Electricity Authority (CEA) \cite{central_electricity_authority_mapping_2020} and Council of Energy, Environment, and Water (CEEW) respectively \cite{karthik_ganesan_coal_2021}. These two databases have the operating SHR values from measurements, as well as technical features such as the boiler design, age as of 2020, power capacity, and plant ownership, but they only cover part of coal plant capacity in India. Compared to the CEA database, the CEEW database has the most updated SHR records in 2022 before the COVID-19 pandemic. However, no database comprehensively covers the operating SHR for all Indian coal plants. This gap motivates the creation of a well-documented, open-access SHR database for Indian coal plants using ML prediction techniques to supplement missing data.

\begin{figure}[!h]
          \centering
         \includegraphics[width=3.7in]{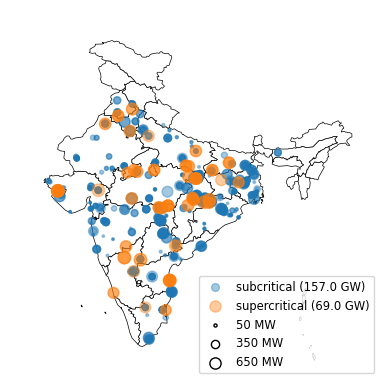}
        \caption{Geographical locations of the operating 806 Indian coal plant units in the research scope. The existing coal-fired power capacity consists of 157 GW from 704 subcritical units and 69 GW from 102 supercritical units}
         \label{coal_locations}
\end{figure}

 In this data descriptor, we present a dataset of predicted SHRs for 806 Indian coal plants extended from previous works and presenting the most comprehensive coverage to date. Figure \ref{coal_locations} shows the locations of all the operating coal plants in India based on the GEM database \cite{global_energy_monitor_global_2024}. We filtered out coal plant units using ultra-supercritical, circulating fluidized bed combustion, or unknown combustion technologies, as these units contribute only 5\% of total power capacity and lack SHR measurement data of their boiler designs. The remaining 806 coal plants are included in our dataset, covering 226 GW in total -- 157 GW from 704 subcritical units and 69.2 GW from 102 supercritical units.

\section*{Methods}

\begin{figure}[!h]
          \centering
         \includegraphics[width=6in]{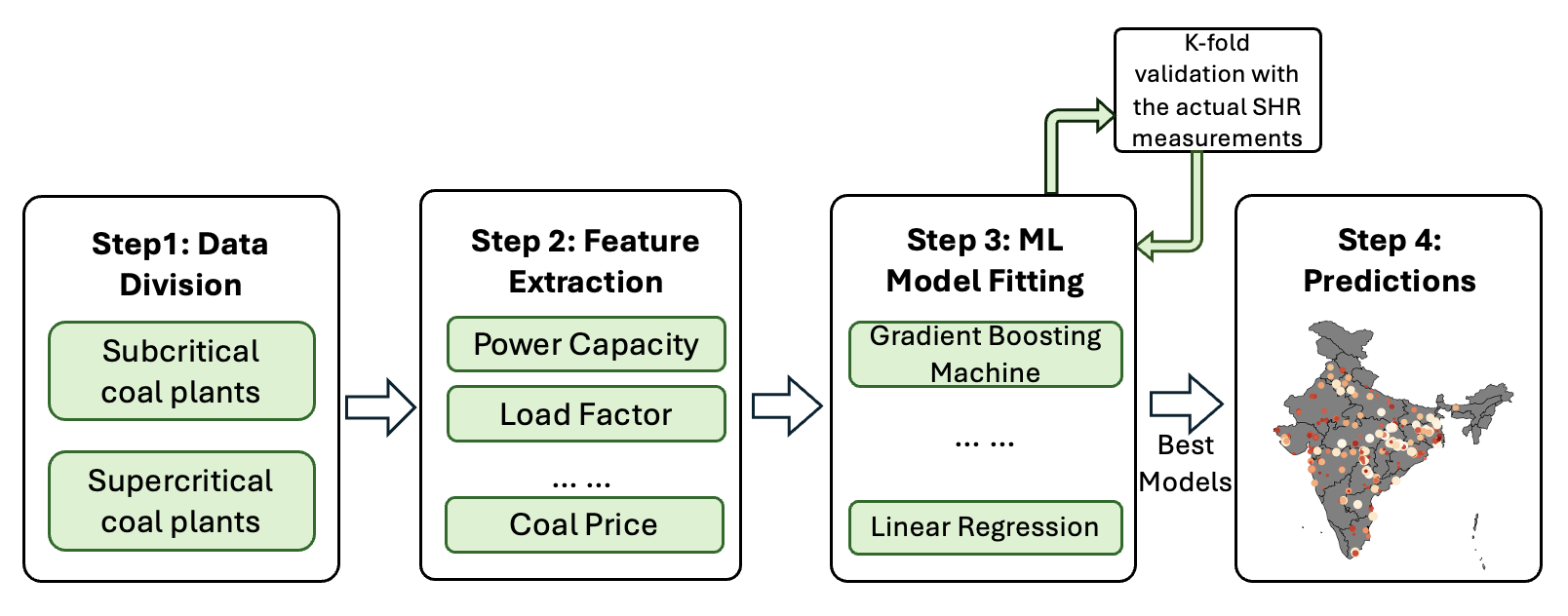}
        \caption{Four steps to predict the operating SHR using ML models for the database}
         \label{Flow_chart}
\end{figure}

Figure \ref{Flow_chart} illustrates the four steps to predict the operating SHR using ML models for the database. Since the boiler design directly determines the SHR of coal plants, we first categorized coal plant datasets into two groups, subcritical and supercritical units. Then, we extracted a variety of prediction features and leveraged the SHR measurements from the CEEW database to train several ML models. Thirdly, we selected the best model for each group based on the prediction performances in the k-fold cross-validation. Finally, we predicted the SHR value for 806 coal plant units based on their records in the GEM database using the trained ML models specific to their boiler design.

\subsection*{Feature Extraction}

To predict SHR, we selected a range of features covering technical parameters, environmental factors, and geographical locations. Specifically, we included: power capacity, plant age, load factor, water stress, coal price, and power system regions. Prior to modeling, we scaled all predictor variables to values between 0 to 1 using the Min-Max scaling method, ensuring consistent variable ranges. By accounting for this diverse set of influential factors, our models aim to capture the complex interplay of technical, environmental, and spatial determinants of SHR across India's coal plant fleet.

\begin{figure}[!h]
          \centering
         \includegraphics[width=6in]{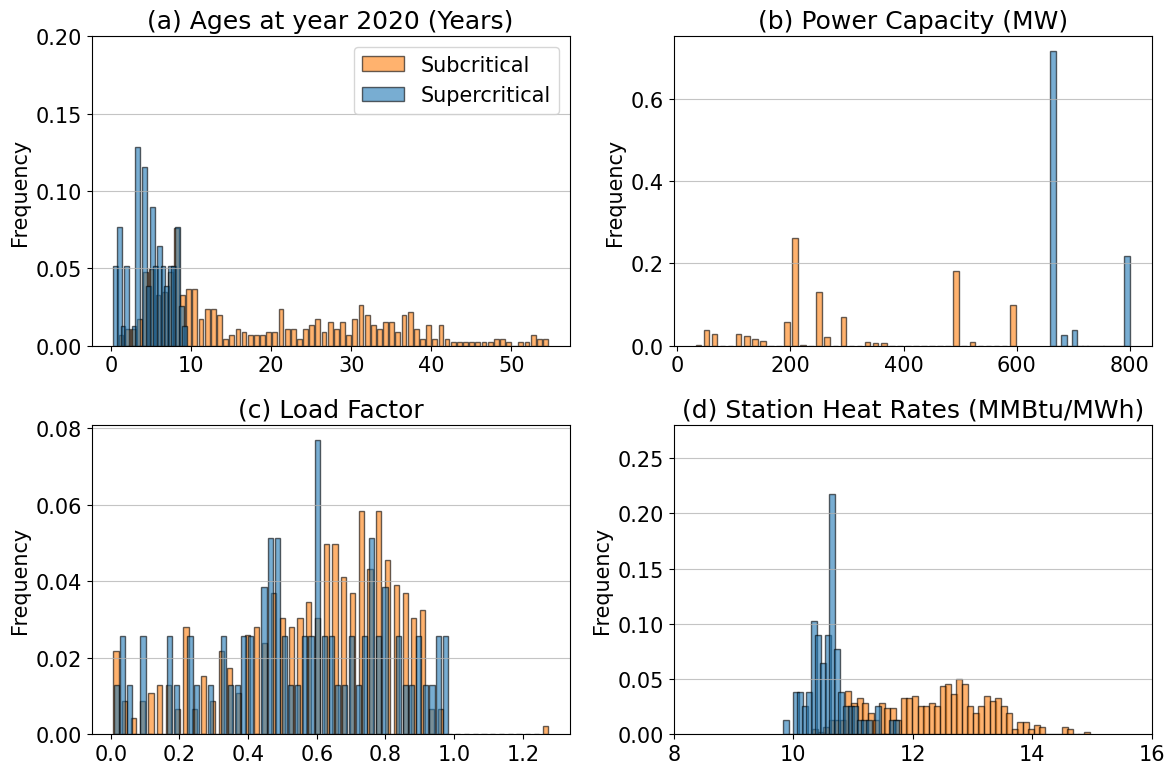}
        \caption{Feature distributions of the unit-level characteristics (a) Plant ages at year 2020 (years) (b) Power capacity (MW) (c) Average load factor and (d) Station heat rate (MMBtu/MWh); The orange bars represent distributions of subcritical coal plants, and the blue bars represent distributions of supercritical coal plants \cite{karthik_ganesan_coal_2021}}
         \label{coal_characters}
\end{figure}

The technical features include the age, load factor, and power capacity of each coal-fired power unit.  Figure \ref{coal_characters} shows these feature distributions from the CEEW database \cite{karthik_ganesan_coal_2021}. Feature distributions of subcritial and supercritical units have distinctive differences. The median age of supercritical units is around five years and much less than the median age of subcritical units. The SHR values for subcritical units exhibit a relatively wider distribution, with a considerable number of plants displaying very high heat rates.

\begin{figure}[!h]
          \centering
         \includegraphics[width=6.5in]{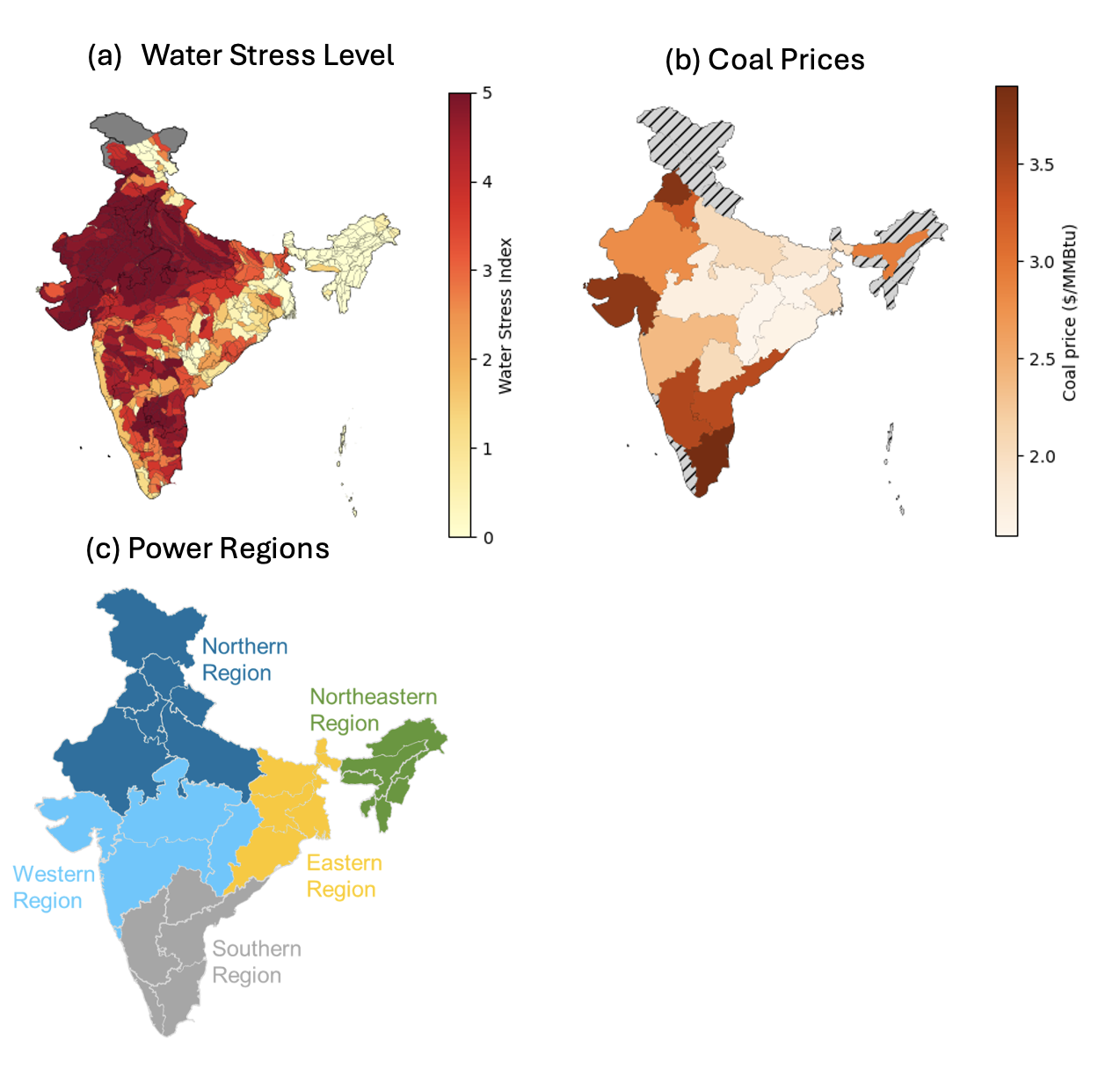}
        \caption{Environmental and geographical prediction features: (a) Water stress level \cite{hofste_aqueduct_2019} (b) Coal prices \cite{karthik_ganesan_coal_2021} and (c) Power system regions \cite{iea_interregional_2020}; The grey or shaded areas in the maps (a) and (b) mean that no data is available or no coal plants are built in the area.}
         \label{water_stress}
\end{figure}

The environmental features include water stress and coal price. The water stress could lead to water shortage and largely impact the cooling systems of coal plants. We utilize the water stress risk index developed by the World Resources Institute \cite{hofste_aqueduct_2019}. The water stress is classified into five levels based on different ranges of water stress index, as listed in Table \ref{Water_stress_level}. As shown in Figure \ref{water_stress} (a), a significant part of India is classified as under "high" or "extremely high" water stress levels. The state-wise coal price is extracted from the CEEW report \cite{karthik_ganesan_coal_2021}, ranging from \$1.59 to \$3.90/MMBtu (Figure \ref{water_stress} (b)). The coal price reflects coal source (domestic or imported) and transportation cost of coal. The eastern regions of India have a much lower coal price than the rest of the country since most of coal mines are located in these regions and, therefore, coal from these regions incur lower transportation costs. Other more distinct geographical features include the power regions of India as a prediction feature. India's power grid is divided into five regions - Northern, Eastern, Western, North Eastern and Southern, as shown in Figure \ref{water_stress} (c) \cite{iea_interregional_2020}. Within each region, thermal power plants are often dispatched based on heuristic rules, which partially impacts the operation and efficiency of coal plants based on location \cite{sengupta_subnational_2022}.

\subsection*{ML Prediction Model Selection and Fitting}
 We build a set of ML models to predict the operating SHR of sub-critical and super-critical units separately. These ML models include: gradient boosting machine (GBM), XGBoost (XGEM), random forest regressor (RF), decision tree regressor (DT), support vector regression (SVR), KNeighbors regressor, ridge regression (RR), and linear regression (LR). These models were chosen due to their diverse underlying algorithms, which provide a comprehensive exploration of potential solutions. \textcolor{black}{The ML model fitting and selection are based on the CEEW database and the additional features as demonstrated in the previous section. Prior to the model fitting, the dataset of 541 coal plants was split into a training set and a test set. The training set consists of 432 of 541 plants (approximately 80\%), while the remaining 109 plants in the test set were used for model validation.}

 We employed k-fold cross validation and grid search to evaluate and tune eight ML models, for predicting SHR values of subcritical and supercritical units. \textcolor{black}{Model performance was assessed using R$^2$ score and three prediction accuracy metrics: mean squared error (MSE), mean absolute error (MAE), and mean absolute percentage error (MAPE). The detailed procedure and results of k-fold cross validation are presented in the Section, Technical Validation. } 
 
 Grid search optimizes the hyper-parameters for each model, and then selects the combination that minimized the average MSE across all splits. The optimal hyper-parameters of eight ML prediction models for subcritical and supercritical units are listed in Tables \ref{hyparameters_sub} and \ref{hyparameters_super}, respectively. Based on the average MSE, GBM predictor and KNeighbors regressor are selected as the best models for sub-critical and supercritical units respectively. Statistical tests, the paired t-test with 5 $\times$ 2 fold cross validation \cite{raschka_model_2018}, are employed to validate the superiority of the selected models. Table \ref{best_models} summarizes the best ML models specific to two boiler designs and evaluation scores through the k-fold validation.

\subsection*{Mapping the Predicted SHR}
Using the trained ML models, we predicted the operating SHR for 806 coal plant units (Figure \ref{coal_locations}) in the GEM database with six input features. A key assumption in our prediction is the average load factor of coal plants. Since the GEM database does not provide load factor information for individual plants, we assumed a uniform load factor of 50\% across the fleet. Our methodology and dataset allow users to predict SHR at varying load factors based on power dispatch results by adjusting this input assumption. This flexibility enables more accurate SHR estimation that aligns with operational regimes of individual plants.

\begin{figure}[!h]
          \centering
         \includegraphics[width=6.5in]{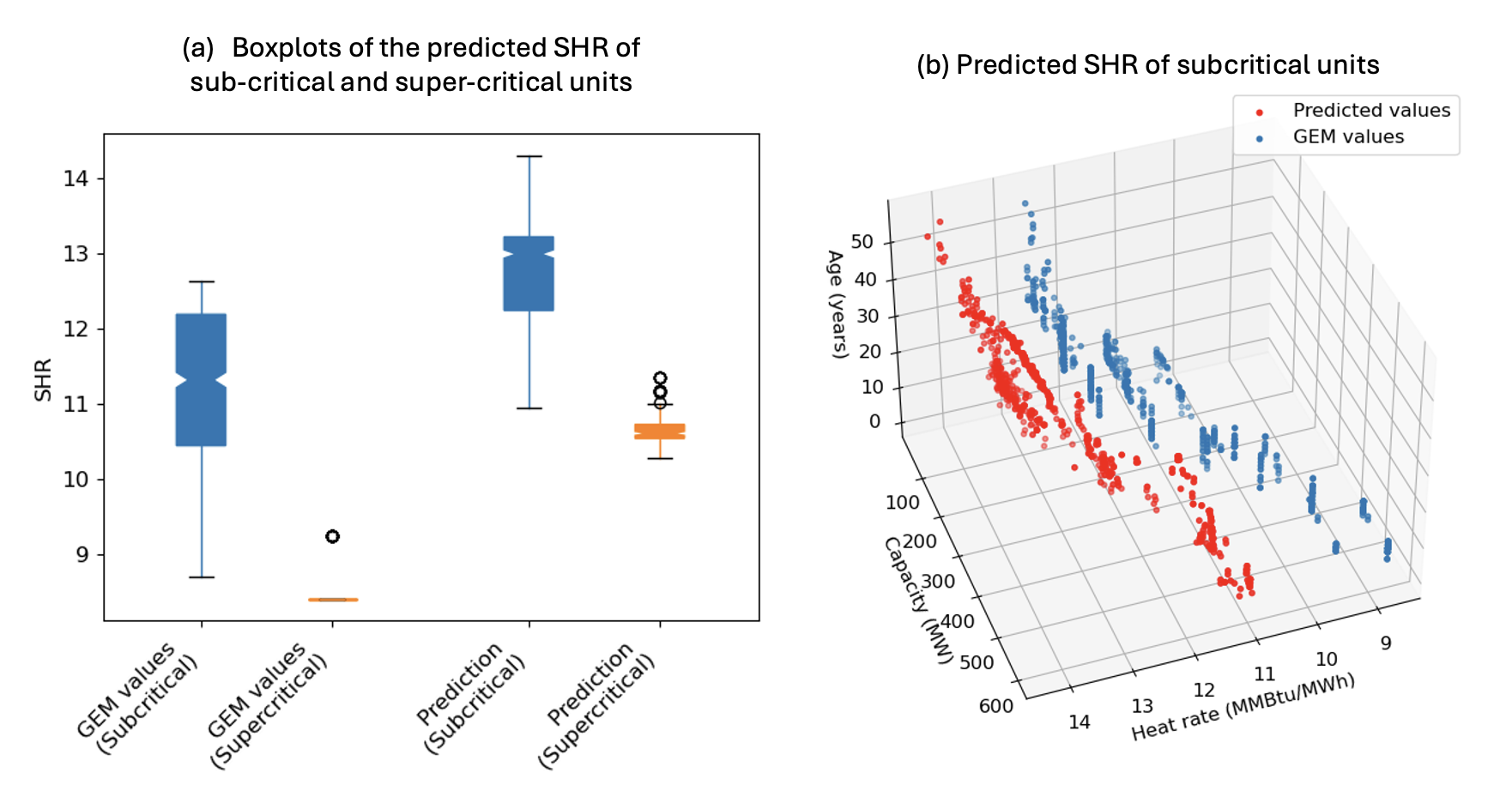}
        \caption{Comparison of the estimated SHR in the GEM database and the predicted SHR from our results: (a) Boxplots for subcritical and supercritical units and (b) Distribution of the predicted SHR for subcritical units with the plant age and power capacity}
         \label{prediction_results}
\end{figure}

\begin{figure}[!h]
          \centering
         \includegraphics[width=6.5in]{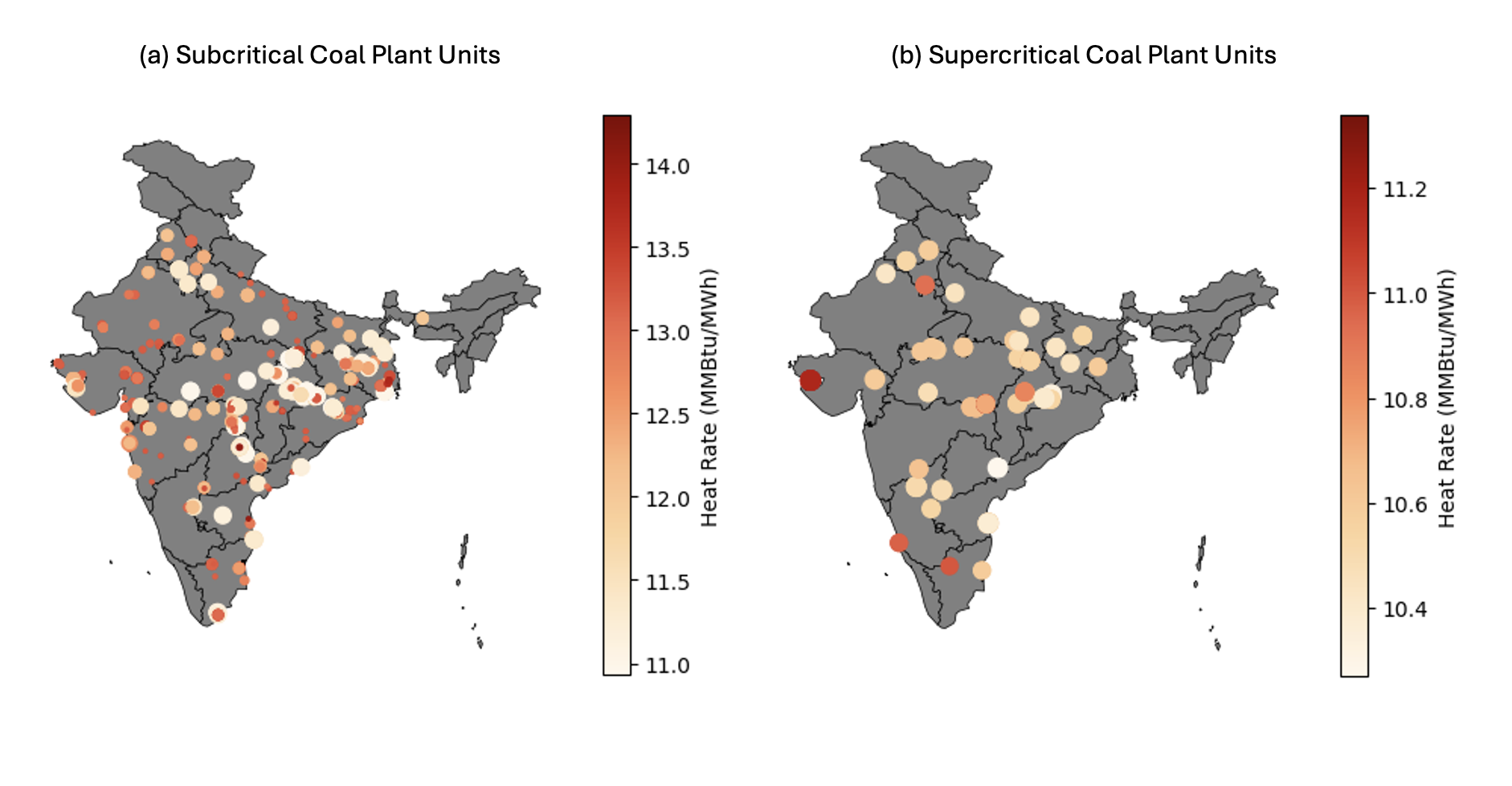}
        \caption{Mapping the predicted SHR of (a) subcritical and (b) supercritical units across India}
         \label{mapping}
\end{figure}

Figure \ref{prediction_results} (a) shows the boxplots to compare the estimated SHR in the GEM database and the predicted SHR from our results for subcritical and supercritical units respectively. The predicted SHR are higher than the estimated values for two unit types, though their distributions regarding age and power capacity are still similar. The average values of the predicted SHR for subcritical and supercritical units are 12.91 MMBtu/MWh and 10.72 MMBtu/MWh, respectively, which are higher than the average value of the estimated SHR of 11.17 MMBtu/MWh and 8.56 MMBtu/MWh. Figure \ref{prediction_results} (b) show distributions of the estimated and predicted SHR considering the effect of the plant age and power capacity. The operating SHR increases as the plant age increases and the power capacity decreases. Figure \ref{mapping} (a) and (b) maps the predicted SHR across India for subcritical and supercritical units respectively.

\subsection*{Feature Importance Analysis}

\begin{figure}[!h]
          \centering
         \includegraphics[width=6in]{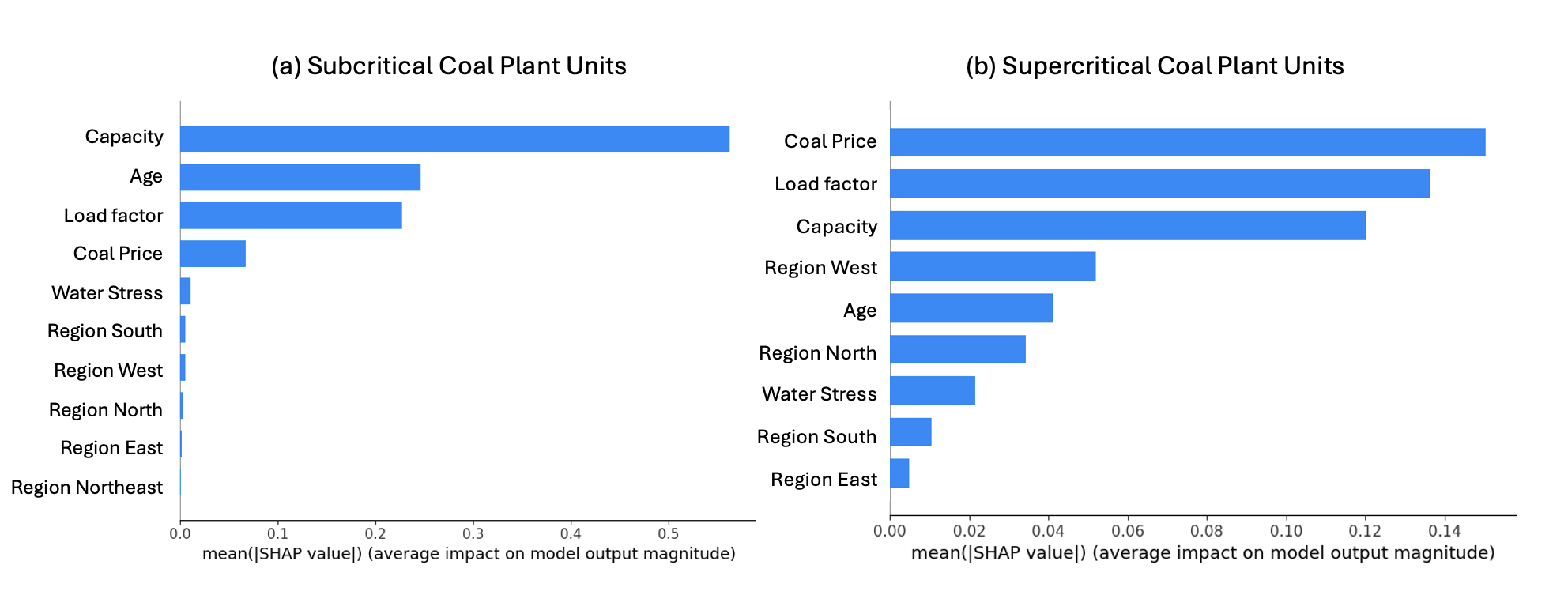}
        \caption{The absolute mean SHAP values for the SHR prediction features for (a) subcritical units and (b) supercritical units}
         \label{shap}
\end{figure}

Finally, we analyzed the feature importance to find the key factors influencing the predicted SHR. We calculated the shapley additive explanations (SHAP) value and ranked the importance of various features based on the absolute mean of the SHAP value. As shown in Figure \ref{shap}, for subcritical units, the three most important features are power capacity, age, and plant load factor. This finding aligns with wide ranges of power capacity and age of subcritical coal plants in India. The coal price is found to be the most important feature for supercritical units, followed by load factor and power capacity. This observation is also evidenced by the similarity between the geographical distribution of SHR in Figure \ref{mapping} (b) and coal price distribution in Figure \ref{water_stress} (b) cross India, which indicates a high coal price could relate to a high SHR value (i.e., low thermal efficiency). Our results show that the plant age does not significantly influence the SHR of supercritical units, due to the relatively young age profile of supercritical coal plants less than 10 years old. Remarkably, the water stress has a limited impact on the SHR prediction for both unit types. This indicates that most Indian coal plants operate independently of local water stress conditions, although their operations are influenced by water shortage.

\section*{Data Records}
\textcolor{black}{All original and predicted data are published in the following Zenodo repository\cite{ding_dataset_2024} (\url{https://zenodo.org/records/10881114}). Table \ref{data_records} provides the detailed descriptions of datasets in the Zenodo repository. The main results can be accessed from \url{gem_predicted_subcritical.csv} and \url{gem_predicted_supercritical.csv}. The following columns (variables) presented in Table \ref{data_headers} include the important features and predicted SHR values. Other columns (variables) are based on the original GEM Indian coal plant database, including locations, names, and owners of coal power plants.}

\section*{Technical Validation}

\begin{figure}[!h]
          \centering
         \includegraphics[width=6.5in]{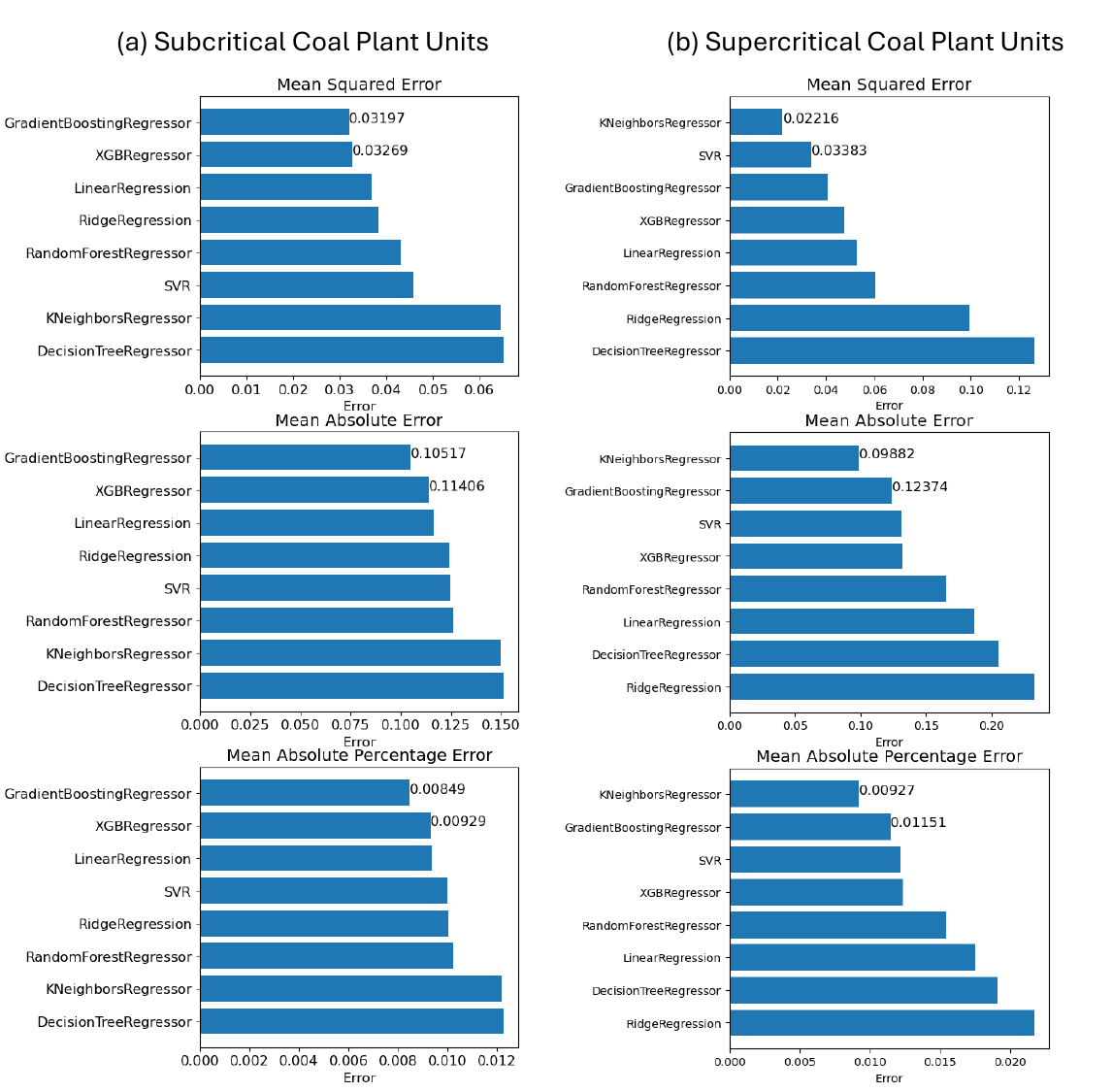}
        \caption{Three evaluation score ranking (MSE, MAE, and MAPE) for the SHR predictions of (a) Subcritical units and (b) Supercritical units. The scores of the top two best models are labeled to the right of the bar.}
         \label{evaluation_score}
\end{figure}

\begin{figure}[!h]
          \centering
         \includegraphics[width=6.5in]{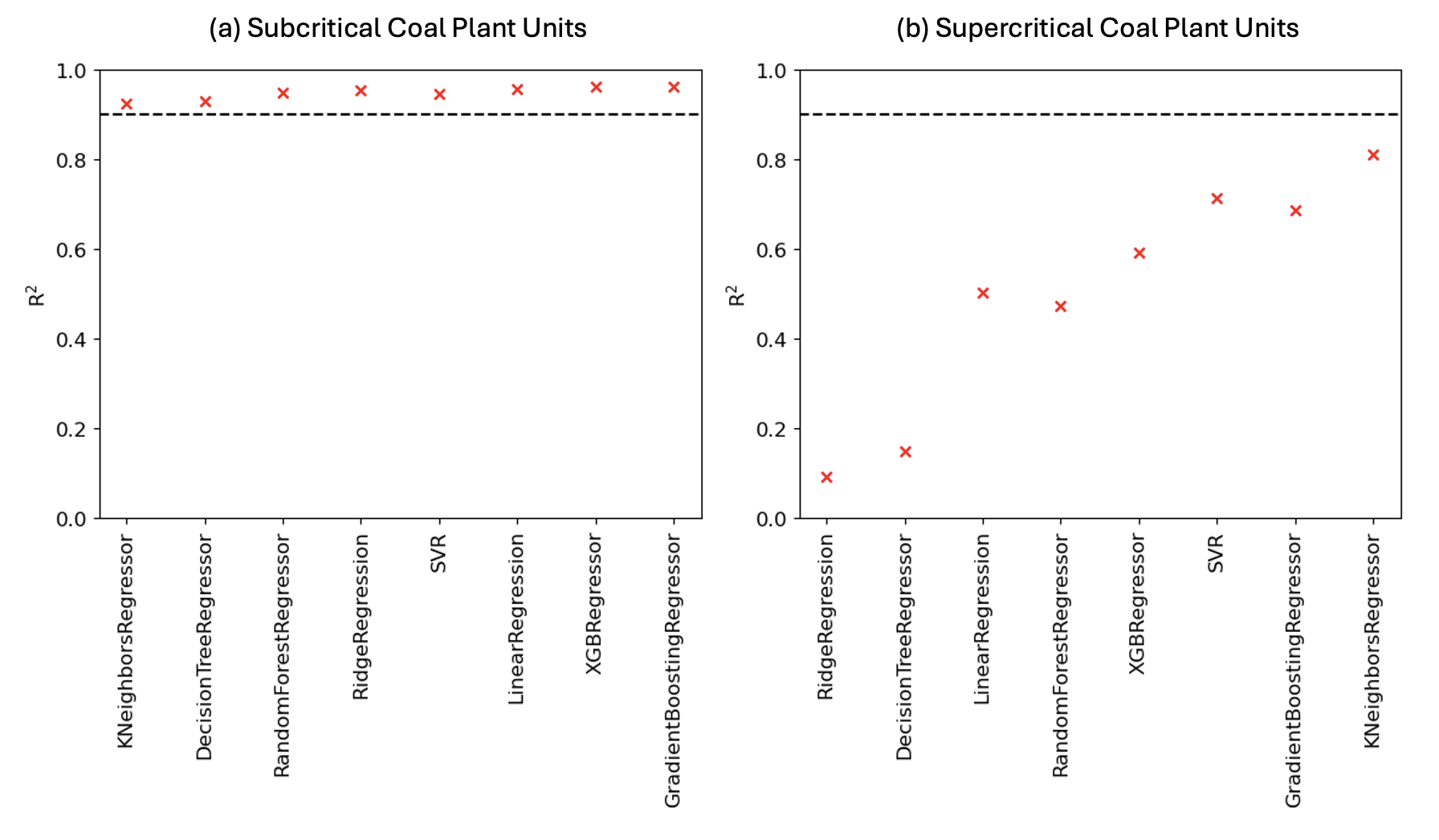}
        \caption{\textcolor{black}{R$^2$ scores for the operating SHR predictions of (a) subcritical and (b) supercritical coal plant units. The dashed line presents R$^2$ = 0.9}}
         \label{R_square_score}
\end{figure}

 The k-fold cross-validation (k=5) and grid search techniques are performed on eight ML models to guarantee the prediction accuracy. For the k-fold cross-validation, the datasets for subcritical and supercritical units were each divided into five folds, with each fold served as the validation set once, while the remaining four folds formed the training set. Figure \ref{evaluation_score} shows three accuracy evaluation score rankings (MSE, MAE, and MAPE) of the SHR predictions for the two unit types respectively. The scores of the top two best models are labeled to the right of the bars. \textcolor{black}{Figure \ref{R_square_score} shows R$^2$ scores of the SHR predictions for the two unit types. The R$^2$ score presents the goodness of fit of different prediction models. The values are usually between 0 to 1. The R$^2$ score of 1 means a perfectly fitted model, and the R$^2$ score of 0 or below means that there is no added value for using the model compared to using the average of the data set.}

 \textcolor{black}{For sub-critical coal power units, the R$^2$ scores of all the models are above 0.9, meaning that all ML models are well-fitted. For super-critical coal power units, the highest R$^2$ score value is 0.8 for the k-Neighbors regression model, which is selected as the best model. These R$^2$ scores also demonstrate the sample sufficiency for training the selected ML models. Several ML models, such as Decision Tree and Random Forest, have a low R$^2$ score below 0.5. This is mainly due to the limited data points used to develop the tree depth.}

\textcolor{black}{To further validate the superiority of the selected model and avoid statistical randomness, we introduce the paired t-test with the 5 $\times$ 2 cross-validation \cite{raschka_model_2018} to compare the first and second-best ML models chosen for the SHR predictions. The dataset is split into 50\% training and 50\% testing sets. We fit two prediction models using the training split in each cross-validation iteration and compare their performances. Then, we rotate the training and test sets and repeat the validation. Under the null hypothesis that two models have equal performance, we compute the t statistics and find the p-value at five degrees of freedom (n=k=5). As summarized in Table \ref{best_models}, the t-values for the SHR predictions of subcritical and supercritical coal plants are 2.97 and 3.20, respectively, resulting in a p-value less than 0.05. We therefore conclude that the two prediction models have significant differences, and the selected first-best ML model is better.}

\section*{Usage Notes}

 \textcolor{black}{All the datasets, code, and environmental dependencies are programmed in Python and provided as an open-accessed source. We predict the most up-to-date number of coal plant units in India (806 units in Jan 2024), and built a visualization platform for dataset download \cite{wong_visulization_2024} (\url{https://states-mode--admirable-creponne-5225d2.netlify.app/}). Our approach can be applied to new coal plant units in the future. When applying new datasets by following the four steps in Figure \ref{Flow_chart}, there are a few important considerations. Firstly, our training dataset does not consider the novel combustion technology (e.g., ultra-supercritical) or retrofitting technology (e.g., Flue Gas Desulfurization module \cite{us__international_trade_administration_india_2020}), as these coal plants only contribute to less than 8\% of the current total coal plant power capacity. Second, the R$^2$ score evaluation should be performed to validate the data sufficiency of new samples and model fitness before comparing the model accuracy. Third, the search ranges for hyperparameters of various ML models (Step 3) might require adjustments depending on the dataset.}

\section*{Code availability}

The code used to generate features and predict SHR database is available within the following GitHub repository \cite{ding_code_2024} (\url{https://github.com/yifueve/India_SHR_dataset}). Detailed code descriptions can be found in Table \ref{code_records}. 



\section*{Acknowledgements}

This work was supported by IHI Corporation, Japan. Jansen Wong was supported via MIT Undergraduate Research Opportunities Program (UROP).

\section*{Author contributions statement}

Y.D. conceptualized the method. Y.D. and J.W. developed the models, generated the data, validated the output, and completed the initial draft. S.P. contributed to collecting the original coal plant data, developing methods and reviewing the draft. D.M., G.Z, and R.S. contributed to the  funding acquisition, project supervision and reviewing the draft.

\section*{Competing interests}

The authors declare no competing interests.

\section*{Figures \& Tables}

\begin{table}[h!]
\centering
\begin{tabular}{l|c|c|c}\hline
Sources & Date & Coal-fired Units / Capacity Coverage & SHR \\\hline
GEM \cite{global_energy_monitor_global_2024} & 2024 & 840 Coal Plant Units / 234 GW & Estimations\\
CEA \cite{central_electricity_authority_mapping_2020} & 2009 & 85 Coal Plant Units / 16.7 GW & Measurements \\
CEEW  \cite{karthik_ganesan_coal_2021} & 2022 & 541 Coal Plant Units / 194 GW & Measurements\\\hline
\end{tabular}
\caption{\label{Databases} Existing databases of India coal plants}
\end{table}

\begin{table}[h!]
\centering
\begin{tabular}{l|c}\hline
Water Stress Level & Water Stress Index \\\hline
Low & $<1$ \\
Low to medium & 1-2 \\
Medium to high  & 2-3 \\
High  & 3-4\\
Extremely high  & $>$ 4\\ \hline
\end{tabular}
\caption{\label{Water_stress_level} Water Stress Index and Classifications}
\end{table}

\begin{table}[h!]
\centering
\begin{tabular}{l|c|c|c|c|c|c}\hline
Unit Types & Best ML Models & MSE & MAE & MAPE & R$^2$ score & p-value\\\hline
Subcritical & GBM & 0.032  & 0.105  & 0.849 \% & 0.810 & < 0.05\\\hline
Supercritical & KNeighbors & 0.022  & 0.099  & 0.927 \% & 0.967 & < 0.05 \\\hline
\end{tabular}
\caption{\label{best_models} \textcolor{black}{The best ML models specific to two unit types with prediction accuracy scores, R$^2$ scores and p-value from the statistical test when compared with the second-best model}}
\end{table}

\begin{table}[h!]
\centering
\begin{tabular}{c|l|l}\hline
No. & Descriptions & File names  \\\hline
1 & WRI water stress geodata\cite{hofste_aqueduct_2019} & \url{indiawater.geojson}\\
2 & State-wise water stress \cite{hofste_aqueduct_2019}  & \url{State_water_stress.csv} \\
3 & State-wise coal price \cite{karthik_ganesan_coal_2021} & \url{State_wise_coal_price.csv} \\
4 & Indian power system regions and states \cite{iea_interregional_2020} & \url{30_to_5zones.csv}\\
5 & The official map of India \cite{indian_remote_sensing_and_gis_india_2017} & \url{india-polygon.shp}\\
6 & CEEW data with additional features (subcritical) \cite{ding_dataset_2024} & 
\url{CEEW_subcritical_with_ws_price.csv} \\
7 & CEEW data with additional features (supercritical) \cite{ding_dataset_2024} & \url{CEEW_supercritical_with_ws_price.csv} \\
8 & GEM data for all Indian coal plants \cite{global_energy_monitor_global_2024} & \url{India_coal_power_plants.csv} \\
9 & GEM data with additional features \cite{ding_dataset_2024} & \url{gem_with_ws_price.csv} \\
10 & GEM data with the predicted SHR (subcritical) \cite{ding_dataset_2024} & \url{gem_predicted_subcritical.csv} \\
11 & GEM data with the predicted SHR (supercritical) \cite{ding_dataset_2024} & \url{gem_predicted_supercritical.csv} \\
\hline
\end{tabular}
\caption{\label{data_records} Data Records of this Research}
\end{table}

\begin{table}[h!]
\centering
\begin{tabular}{l|l|p{8cm}}\hline
Column headers (Variables) & Units & File names  \\\hline
Capacity & MW & Nameplate power capacity (MW) of coal power plants\\
bws\_score & - & WRI water stress index \\
Coal price & \$/MMBtu & Coal price based on 30 Indian states \\
Remaining power plant lifetime & Years & The remaining power plant lifetime given a 50-year lifetime\\
Combustion technology & - & Subcritical or supercritical coal plants \\
Heat Rate & Btu/kWh & 
Calculated SHR values of subcritical and supercritical coal plants in the GEM database \\
Predicted\_HR & MMBtu/MWh & 
Predicted operating SHR values of subcritical and supercritical coal power plants \\
\hline
\end{tabular}
\caption{\label{data_headers} Output data headers descriptor}
\end{table}

\begin{table}[h!]
\centering
\begin{tabular}{c|l}\hline
ML models & Optimal Hyper-parameters  \\\hline
GBM & 'learning rate': 0.1, 'loss': 'absolute error', 'max depth': 15, 'number of estimators': 200 \\
XGBM & 'learning rate': 0.1, 'max depth': 5, 'no of estimators': 400 \\
RF & 'min samples split': 2, 'min samples leaf' = 1, 'max depth': None \\
DT & 'min samples split': 2, 'min samples leaf' = 1, 'max depth': None\\
SVR & 'C': 0.1, 'gamma': 0.01, 'kernel': 'linear'  \\
kNeighbors & 'no of neighbors': 2, 'weights': 'distance' \\
RR & 'alpha'=1.0 \\
LR & - \\
\hline
\end{tabular}
\caption{\label{hyparameters_sub} The optimal hyper-parameters for eight ML models (Subcritical units)}
\end{table}

\begin{table}[h!]
\centering
\begin{tabular}{c|l}\hline
ML models & Optimal Hyper-parameters  \\\hline
GBM & 'learning rate': 0.1, 'loss': 'absolute error', 'max depth': 30, 'number of estimators': 300 \\
XGBM & 'learning rate': 0.1, 'max depth': 2, 'no of estimators': 600 \\
RF & 'min samples split': 3, 'min samples leaf' = 2, 'max depth': None \\
DT & 'min samples split': 3, 'min samples leaf' = 2, 'max depth': None\\
SVR & 'C': 0.01, 'gamma': 1, 'kernel': 'poly'  \\
kNeighbors & 'no of neighbors': 1, 'weights': 'uniform' \\
RR & 'alpha'=10.0 \\
LR & - \\
\hline
\end{tabular}
\caption{\label{hyparameters_super} The optimal hyper-parameters for eight ML models (Supercritical units)}
\end{table}

\begin{table}[h!]
\centering
\begin{tabular}{c|l|l}\hline
No. & Descriptions & File names  \\\hline
1 & Prediction framework for subcritical units & \url{run_models_subcritical.ipynb} \\
2 & Prediction framework for supercritical units & \url{run_models_supercritical.ipynb} \\
3 & Code for mapping the water stress levels & \url{india_water_geojson.ipynb} \\
4 & Code for mapping the SHR prediction results & \url{india_map.ipynb} \\
5 & Hyperparameters tuning for predictions (subcritical)   & \url{GRID_search_subcrit.ipynb} \\
6 & Hyperparameters tuning for predictions (supercritical)   & \url{GRID_search_supercrit.ipynb} \\
7 & \textcolor{black}{Statistical test for the best two prediction models (subcritical)}   & \url{statistical_test_subcritical.ipynb} \\
8 & \textcolor{black}{Statistical test for the best two prediction models (supercritical)}   & \url{statistical_test_supercritical.ipynb} \\
9 & Environmental dependencies of the code & \url{enviormental.yml} \\
\hline
\end{tabular}
\caption{\label{code_records} Code to generate the database}
\end{table}
 
\end{document}